\documentstyle[twoside,fleqn,knuckle]{article}
\newcommand{\ncm}{\newcommand}
\ncm{\be}{\begin{equation}}
\ncm{\ee}{\end{equation}}
\ncm{\bea}{\begin{eqnarray}}
\ncm{\eea}{\end{eqnarray}}
\ncm{\bedm}{\begin{displaymath}}
\ncm{\eedm}{\end{displaymath}}
\ncm{\dg}{\dagger}
\ncm{\htm}{\hat{\mu}}
\ncm{\dau}{\partial_{\mu}}
\ncm{\dauu}{\partial_{\nu}}
\ncm{\Dl}{\Delta}
\ncm{\bt}{\beta}
\ncm{\ta}{\tau}
\ncm{\sg}{\sigma}
\ncm{\eps}{\varepsilon}
\ncm{\nn}{\nonumber}
\ncm{\ra}{\rightarrow}
\ncm{\bts}{\beta_{\mbox{\small s}}}
\ncm{\btg}{\beta_{\mbox{\small g}}}
\input{psfig}
\title{Gribov Copies And Other Gauge Fixing Beasties On The
Lattice \\[0.4\baselineskip]
\vspace*{-25mm}
{\normalsize\noindent November 1992 \hfill LTH 291 \\
\mbox{} \hfill hep-lat/9211018 \\}
\vspace*{17mm}
}
\author{{\large Arjan Hulsebos}\thanks{email: {\tt arjanh@castle.ed.ac.uk}}
\\[0.5\baselineskip]
DAMTP, Chadwick Tower, University of Liverpool, P.~O.~Box 147,
Liverpool, L69 3BX, UK. }
\begin{document}
\begin{abstract}
We study the nature of gauge fixing ambiguities in two dimensional gauge
theories. We find that these ambiguities can be related to the asociated spin
model. They can be eliminated by means of a (multigrid) annealing algorithm.
\vspace*{-1\baselineskip}
\end{abstract}
\maketitle

Gauge fixing is by now a useful tool in lattice gauge
calculations. Smooth gauges, like the Coulomb gauge and the
Landau gauge, are commonly used in propagator calculations
\cite{Berna90}, as well as in wavefunction calculations
\cite{wavefunctions}. In monopole studies, the so-called
maximally Abelian gauge is frequently employed \cite{Suzu92}.
However, since the work of Gribov \cite{Grib78}, it is known that
`fixing the gauge is not quite fixing the gauge'. In other words,
it is possible to {\it locally}\/ fix the gauge, but it need not
always be possible to {\it globally}\/ fix the gauge. Since one
uses a slightly stronger condition on the lattice than in the
continuum, it was long hoped that the Gribov ambiguity was absent
on the lattice. A few years ago, it was shown that these hopes
were idle \cite{copies}. In order to get a better understanding
of the nature of the Gribov ambiguity on the lattice, we shall
study the occurrence of copies in 2-d U(1) and 2-d SU(2) when
fixing to the Landau gauge.

The continuum version of the Landau gauge condition,
\bedm
\dau A^g_{\mu} = 0,\;\; A^{g}_{\mu} = g^{\dg}(\dau \,+\, A_{\mu})\,g,
\eedm
translates to the lattice as
\bea
\mbox{Im} \sum_{\mu}\, [U^g_{\mu}(x) -
U^{g\,\dg}_{\mu}(x-\htm)]_{\mbox{\small traceless}} = 0, & &
\label{gaugecondition}  \\
U^g_{\mu}(x) = g^{\dg}(x)U_{\mu}(x)g(x+\htm).
\;\;\;\;\;\;\;\;\;\;\;\;\;\;\;\;\;\; & & \nn
\eea
As was mentioned above, usually a slightly stronger condition is
used, namely minimizing a functional $F$, where $F$ is given by
\be
F(g,U) = \frac{1}{dVN} \sum_{x,\mu} \,\mbox{Re Tr}\,[1\!\!1\, -
U_{\mu}^g(x)], \label{latticeLandau}
\ee
for SU(N) in $d$ dimensions on a lattice containing $V$ lattice
sites. It is easy to see that if (\ref{latticeLandau}) is
minimized for a given gauge configuration $U$,
(\ref{gaugecondition}) is automatically satisfied.

The Gribov problem, finding multiple solutions for the Landau
gauge fixing condition
\bedm
\dau A_{\mu} + [D_{\mu}(A)\,,\,(\dau g)g^{\dg}] = 0,
\eedm
$D_{\mu}(A)$ being the covariant derivative, is equivalent to
finding local minima for the functional $F$.

\begin{figure}
\vspace*{-2.2cm}
\hspace*{-1.0cm}
\centerline{%
    \hbox{%
    \psfig{figure=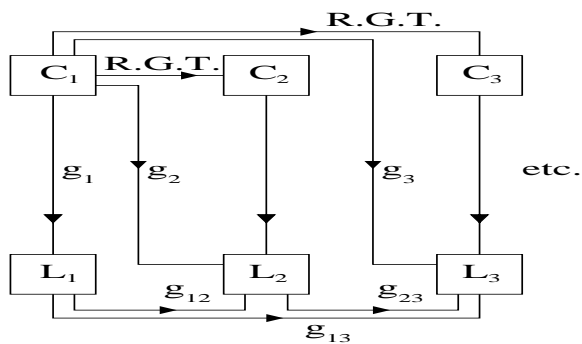,height=3in,width=5in}%
   }%
}
\vspace{-1.7cm}
\caption{The search for Gribov copies. R.G.T.\ stands for random gauge
transformations}
\label{GCS}
\vspace*{-0.5cm}
\end{figure}

The search for Gribov copies on the lattice is done as follows.
We start from an initial configuration $C_1$, and fix that
configuration to the Landau gauge, keeping track of the gauge
transformations involved. The resulting configuration is called
$L_1$. We then perform a random gauge transformation on $C_1$, to
obtain $C_2$, and fix $C_2$ to the Landau gauge $L_2$. We can
repeat this as many times as we want. Since we have kept track of
the gauge transformations $g_i$ between $C_1$ and $L_i$,
we can construct the gauge transformation $g_{ij}$
between $L_i$ and $L_j$:
\be
g_{ij}(x) = g^{\dg}_i(x)\,g_j(x). \label{Gribtrans}
\ee
This procedure is displayed in figure \ref{GCS}.

One can look upon the gauge fixing procedure from another point
of view. Minimizing the functional $F$ can be seen as finding the
ground state of a spin system with pseudo-random couplings $U$,
depending on both position and direction, with an action given by
\be
S_U(g) = -\bts\,F(g,U). \label{gf-action}
\ee
We have put `pseudo-random couplings $U$', since only for gauge
field coupling constant $\btg=0$ are the
couplings $U$ truly random. For finite $\btg$,
the $U$ are constrained to form plaquettes $U_{\mbox{\small P}}$,
which are to be taken from an equilibrium ensemble.

Making a random gauge transformation is, in some sense, creating
a spin configuration at spin coupling constant $\bts = 0$.
The standard procedure to find the minimum of
(\ref{latticeLandau}) is employing a relaxation algorithm. Albeit
impractical, this can be `simulated'\/ by performing Monte Carlo
at $\bts = \infty$. We should therefore not be
surprised if defects get trapped in the final, gauge fixed
configuration: we don't expect a perfect crystal to be formed by
dropping liquid iron into liquid helium.

To determine whether we have reached convergence, we would like
to use $F$. However, for non-trivial gauge configurations we do
not know the the minimal value(s) of $F$. Therefore, we define
\bedm
f_i = F_{i-1} - F_i.
\eedm
$F_i$ is the value of $F$ after $i$ iterations. We can
assign a relaxation time $\ta$ to the minimization process by
defining $\ta$ by
\be
f_i = C \exp(-i/ \ta).  \label{tau}
\ee

For 2-d U(1), the asociated spin model is the well known
XY-model. It possesses a Kosterlitz-Thouless phase transition at
a finite coupling $\bt_c$. The system is described by vortices
and antivortices for couplings $\bt < \bt_c$, and by spin waves
for couplings $\bt > \bt_c$.

\begin{figure}
\vspace{-2.3cm}
\centering
\centerline{%
   \hspace{9.5mm}%
   \hbox{%
   \psfig{figure=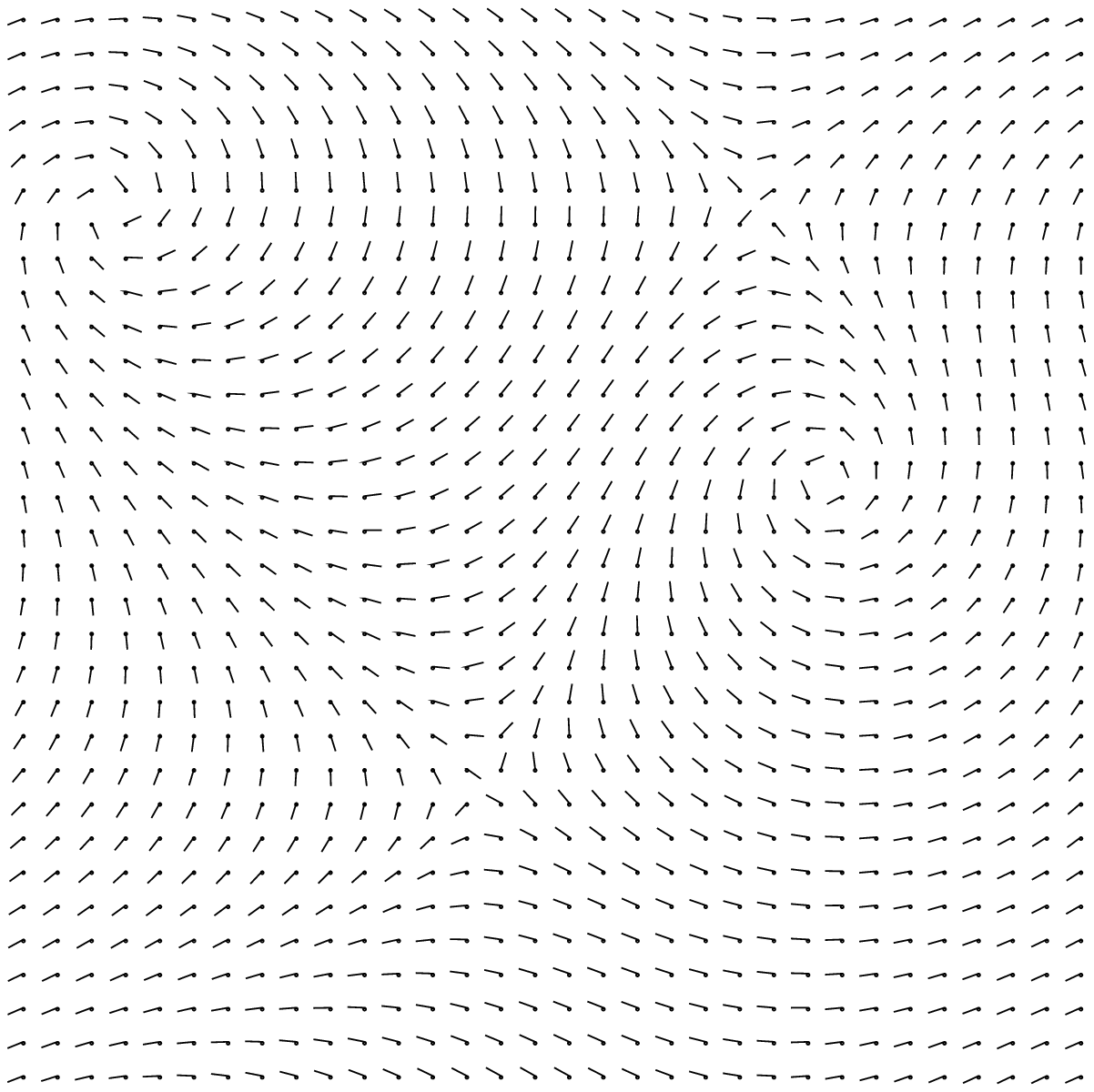,height=4.3in,width=4.3in}%
       }%
}%
\vspace{-2.5cm}
\caption{The final, Landau gauge fixed configuration, starting
from pure gauge on a $32^2$ lattice. A few, widely seperated
vortices and antivortices, as well as two spin waves, are left.}
\label{spinconf}
\vspace*{-.5cm}
\end{figure}

We created pure gauge configurations by taking $U_{\mu}(x) =
g^{\dg}(x)g(x+\htm)$, and fixed these to the Landau gauge. We
used a standard relaxation algorithm, combined with multigrid
\cite{multifix}. We experienced, even on a $16^2$ lattice,
difficulties in reaching the true Landau gauge, $U_{\mu}(x) = 1$.
On larger lattices it became virtually impossible to reach the
true Landau gauge. Figure \ref{spinconf} displays a typical gauge
fixed configuration for a pure gauge configuration on a $32^2$
lattice.

Although the inclusion of overrelaxation in the minimization
algorithm improves on the relaxation time $\ta$ and is able to
remove all the vortex-antivortex pairs on the smaller
lattices, the combined algorithm is not able to eradicate all
pairs on the larger lattices.  Nevertheless, in the following we
shall make use of the combined algorithm.

Turning to gauge fields taken out of an equilibrium ensemble, we
find basically the same results. We found that the $\ta$ values
depended slightly on the $\bt$ value of the gauge fields, as well
as on the lattice size, but not on the number of (anti)vortices
or spin waves left in the gauge fixed configuration.

\begin{figure}
\vspace{-2.5cm}
\centering
\centerline{%
  \hspace*{-1.3cm}%
   \hbox{%
   \psfig{figure=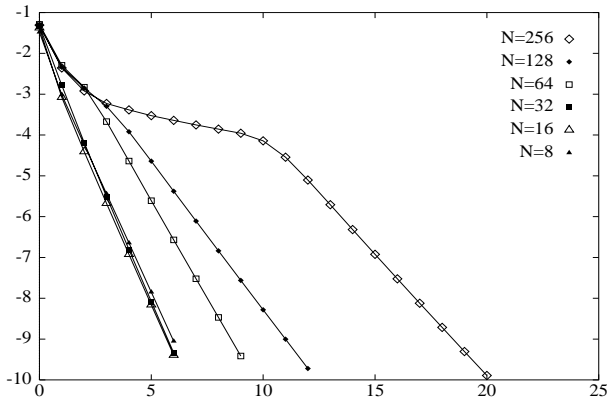,height=3.0in,width=3.3in}%
       }%
}%
\vspace*{-1.3cm}
\caption{$\mbox{}^{10}\log F$ versus the number of
iterations for pure gauge SU(2)
configurations. An overrelaxed minimization algorithm was used in
combination with multigrid with W-cycles. \label{pgsu2}}
\vspace*{-.5cm}
\end{figure}

For 2-d SU(2), the asociated spin model is the O(4) non-linear
sigma model. This model is asymptotically free and does not
contain any topological objects such as instantons or
(anti)vortices. Therefore, we do not expect to find any
obstructions in fixing pure gauge to the Landau gauge. This is
confirmed by figure \ref{pgsu2}.

Taking gauge field configurations from an equilibrium ensemble,
and fixing these to the Landau gauge, we at times do find Gribov copies.

\begin{figure}
\vspace{-2.5cm}
\centering
\centerline{%
  \hspace*{-1.3cm}%
   \hbox{%
   \psfig{figure=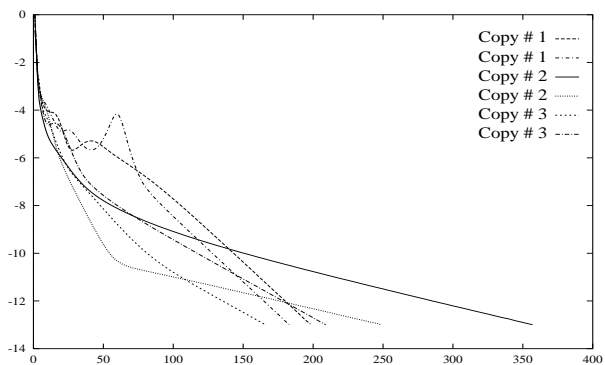,height=3.0in,width=3.3in}%
       }%
}%
\vspace*{-1.3cm}
\caption{$\mbox{}^{10}\log f$ versus the number of
iterations for one SU(2) gauge
configuration. \label{SU2real}}
\vspace*{-.5cm}
\end{figure}

These copies behave differently from the ones we found for 2-d
U(1). The different copies relax with different autocorrelation
times $\ta$ to their minima. This is depicted in figure
\ref{SU2real}.

At fixed lattice size, it is observed that the relaxation time
$\ta$ on average becomes larger when $\btg$ is
increased. Furthermore, the relative occurrence of these copies
becomes smaller; see table \ref{occur}.

\begin{table}
\centering
\begin{tabular}{|c|c||c|c|c|} \hline
N & $\btg$ & $\# = 0$ & $\# =1 $ & $\# \geq 2$ \\ \hline
8  &   3.0 & 26 & 4 & --- \\
   &   6.0 & 27 & 3 & --- \\
\hline
16 &  12.0 & 29 & 1 & --- \\
   &  24.0 & 29 & 1 & --- \\
\hline
32 &  48.0 & 17 & 11 & 2 \\
   &  96.0 & 20 & 7 & 3 \\
\hline
\end{tabular}
\caption{The number of times a gauge configuration yielded $\#$ copies for
several combinations of the lattice size $N$ and the gauge coupling $\bt_g$.
The $\bt_g$ values are chosen to keep the ratio $N/\xi_{\sg}$ roughly constant.
$\xi_{\sg}$ is the string tension correlation length.
\label{occur}}
\vspace*{-0.7cm}
\end{table}

We would like to find an observable $f$ for the gauge transformation
$g_{ij}$ to disinguish between the different Gribov copies.
This observable should be invariant under global spin rotations,
i.~e.\ $f(g_{ij}) = f(g_{ij}h) \;,   \,  h \in $ SU(2), since global
spin rotations correspond to global gauge transformations
$U_{\mu}(x) \ra U^{'}_{\mu}(x) = h^{\dg}U_{\mu}(x)h$.

The first observable to satisfy this criterion is
the most obvious one: the (spin model) energy density
\bedm
e(x) = \sum_{\mu}\,\mbox{Tr}\,
 [g^{\dg}(x)g(x+\htm) + g^{\dg}(x-\htm)g(x)].
\eedm
We use the parametrization
\bedm
g(x) = [g_4 1\!\!1\, - \mbox{i}\, \vec{\sg} \cdot \vec{g}] \,(x),
\eedm
since it allows us to change conveniently
from the O(4) or vector notation, to the
SU(2) or matrix notation, and vise versa. The second observable
is given by
\bea
J(x) & = & \sum_{i} j_{\imath}^2(x), \nn \\
j_{\imath}(x) & = & [ \eps_{\mu \nu} \eps_{ijkl} g_j
\dau g_k \dauu g_l ]\,(x). \nn
\eea

\begin{figure}
\vspace{-3.5cm}
\centering
\centerline{%
  \hspace*{-1.0cm}%
   \hbox{%
   \psfig{figure=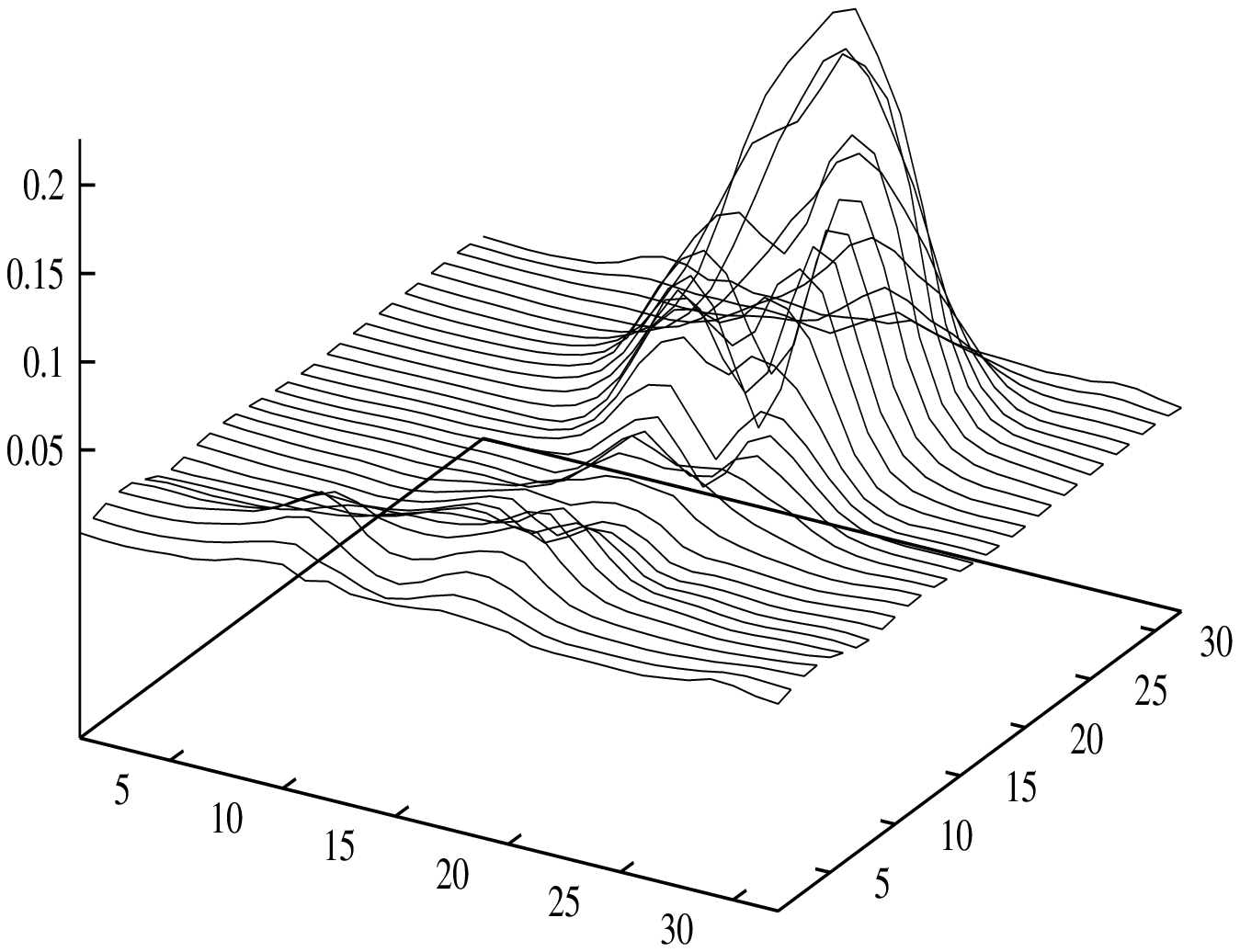,height=3.0in,width=3.3in}%
       }%
}%
\vspace*{-4.0cm}
\centerline{%
  \hspace*{-1.0cm}%
   \hbox{%
   \psfig{figure=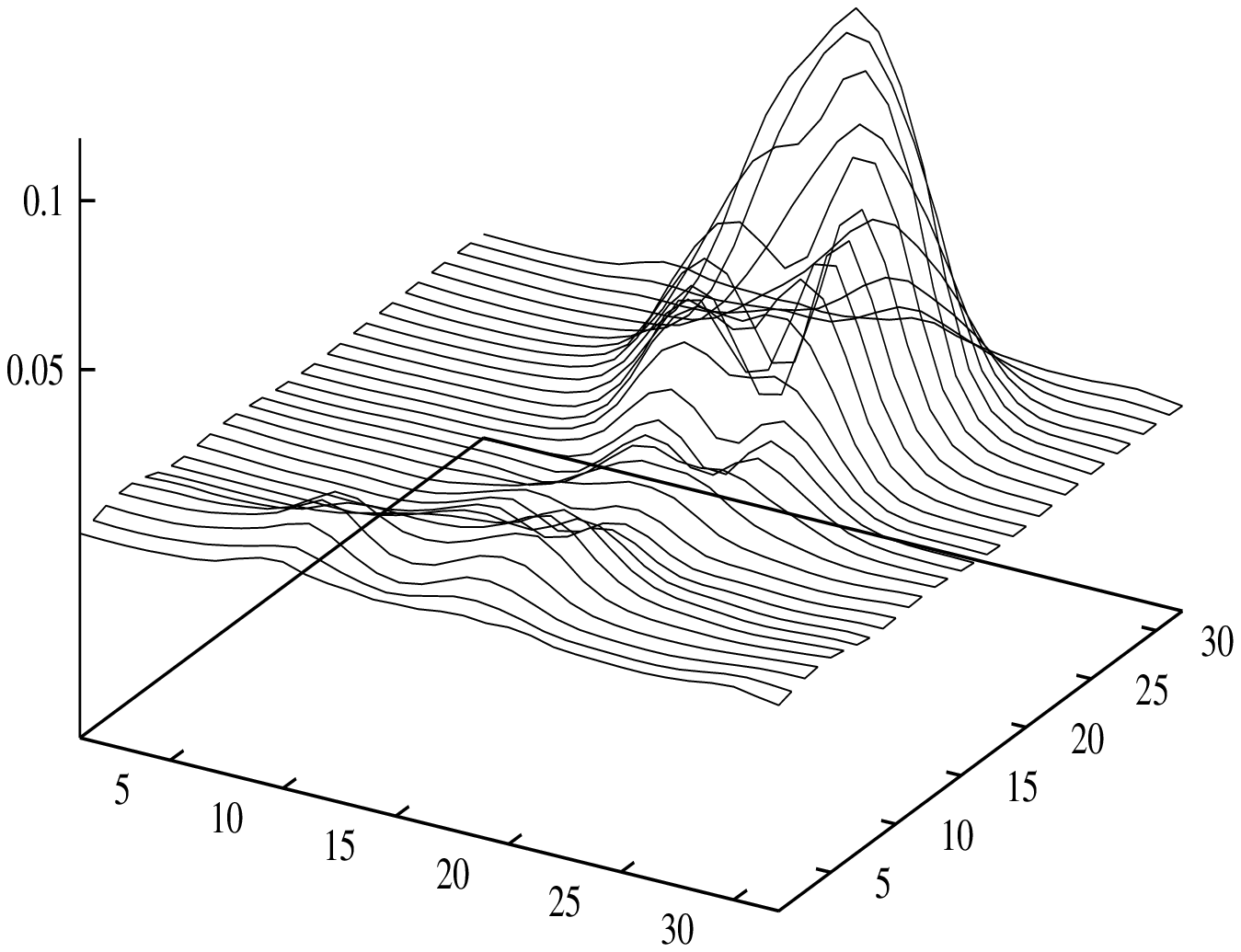,height=3.0in,width=3.3in}%
       }%
}%
\vspace*{-1.3cm}
\caption{ $J$ (top) and $1 - \frac{1}{4} e$ (bottom) for a
gauge transformation
relating a copy to the true Landau gauge, obtained
from a gauge configuration taken from an equilibruim ensemble at $\bt = 48.0$
on a $32^2$ lattice.  \label{copies}}
\vspace*{-1.0cm}
\end{figure}

We have plotted $1 - \frac{1}{4} e(x)$ and $J(x)$ for a
gauge transformation taking one copy
to the absolute minimum; see figure \ref{copies}.
Surprisingly, the two figures are almost identical in shape!
The reason for this similarity is not (yet) understood and is being
studied.

The reason for copies to occur is that once we have come close to a
local minimum, there is no chance to get away from this minimum, due to
the minimization algorithm. This immediately implies a way out: using
an annealing algorithm. The implementation for gauge fixing is clear
from (\ref{gf-action}): we perform Monte Carlo on the spins $g$ using the
action $S_U(g)$ at some value of $\bts$. We should start with a (very)
low value of $\bts$, and gradually increase $\bts$ by small amounts $\Dl\bts$.
After increasing $\bts$, we should let the system reach equilibrium before
increasing $\bts$ once more. When $\bts$ is large enough, we can turn to
ordinary relaxation, with or without overrelaxation. The starting value of
$\bts$ and $\Dl\bts$, as well as the value of $\bts$ where we can turn to
the normal minimization must be determined in practice, and may depend on
$\btg$.

The interpretation of this process is straightforward. Making a random
gauge transformation is (in some sense) taking a spin configuration at
infinite temperature ($\bt \equiv T^{-1}$).  Performing Monte Carlo at low
$\bts$ is nothing more than letting the (spin) system evolve in time, deep
into the high temperature phase. Increasing $\bts$ is then lowering the
temperature, which should be done gradually especially near a phase
transition. If we have been careful enough, we can throw the system into
liquid helium once we are deep into the low temperature phase.

We used this annealing algorithm, using a multigrid algorithm as described
in \cite{multifix} for the Monte Carlo part, for both U(1) and SU(2)
gauge fields. After tuning the aforementioned parameters, we found no copies,
even for very rough gauge fields; $\bt = 24.0$ on a $128^2$ lattice
for SU(2), where $\xi_{\sg}\simeq 4$.

We can draw several conclusions from this finding. Firstly, the copies
for SU(2) are probably a relic from the spin model.

Secondly, the anology with spin glasses, which sometimes is made to explain
gauge fixing ambiguities, may be misleading. To be sure, it strictly holds
for $\btg = 0$ only, since for $\btg \neq 0$ the couplings $U$ are
constrained, as was mentioned before. Especially for larger values of $\btg$,
we should expect that the local minima thin out, and become well seperated.
This is indeed observed, but surprisingly, for $\btg$ values in the range
of $\btg = 6.0$ --- $192.0$ on $32^2$ lattices, we find
$\Dl F_{\infty}/F_{\infty}$ to be in the order of a few percent.
$F_{\infty}$ denotes the final value of $F$ for a given
configuration. This suggests that the deviation from a spin glass starts
at lower values of $\btg$, and therefore, that the problem of finding
the true Landau gauge is not as severe as finding the ground state of
a spin glass. This fact encourages us to apply annealing algorithms
to 4-d gauge fixing problems.
\vspace*{-0.5\baselineskip}

\section*{ACKNOWLEDGEMENTS}

\mbox{} \vspace*{-0.5mm}
We would like to thank Pierre van Baal, Philippe de Forcrand, Jim Hetrick,
Morten Laursen, Chris Michael, Ronald Rietman, Jan Smit and Jeroen Vink for
useful discussions and comments. This work was supported by EC contract SC1
*CT91-0642.
\hspace*{-0.5\baselineskip}

\end{document}